# An *in silico* approach to analyse the influence of carotid haemodynamics on cardiovascular events using 3D tomographic ultrasound and computational fluid dynamics

Sampad Sengupta[1,2,*], Emily Manchester[1], Jie Wang[1], Ajay B. Harish[1], Alistair Revell[1], Steven K. Rogers[3,4,5]

1. Department of Mechanical and Aerospace Engineering, School of Engineering, The University of Manchester, Manchester, UK.
2. Department of Chemical Engineering and Biotechnology, The University of Cambridge, Cambridge, UK.
3. Manchester Academic Vascular Research and Innovation Centre (MAVRIC), Manchester Vascular Centre, Manchester University NHS Foundation Trust, Manchester, UK.
4. Division of Cardiovascular Sciences, School of Medical Sciences, Faculty of Medicine, Biology and Health, University of Manchester, Manchester, UK
5. Division of Nursing, Midwifery and Social Work, School of Health Sciences, Faculty of Medicine, Biology and Health, University of Manchester, Manchester, UK

*Corresponding author: ss3253@cam.ac.uk (Sampad Sengupta)

## Abstract

Analysing the haemodynamics of flow in carotid artery disease serves as a means to better understand the development and progression of associated complex diseases. Carotid artery disease can predispose people to major adverse cardiovascular events. Understanding the nature of carotid blood flow using *in silico* methods enables the extraction of relevant metrics that are not accessible *in vivo*. This study develops computationally efficient means of modelling patient-specific flow, utilising 3D tomographic ultrasound to generate anatomically faithful reconstructions including artherosclerotic plaque, and computational fluid dynamics to simulate flow in these arteries. A computationally efficient model has been proposed here, which has been used to conduct simulations for a large dataset, the results of which where stastitically analysed to test the association of relevant haemodynamic metrics with cardiovascular events. The incidence of major cardiovascular diseases in carotid artery disease patients has been shown to have an association with flow vorticity in the region of interest, and less so with the absolute magnitudes of wall shear stress.

## 1. Introduction

Computational modelling of carotid arteries allows haemodynamic measurement of many parameters which cannot be assessed *in vivo*. Carotid arteries emerge from the aortic arch running along the front of an individual's neck, and are responsible for oxygenating a significant portion of the brain. The common carotid artery (CCA) runs up from the arch and then breaks into the internal carotid artery (ICA) and the external carotid artery (ECA). As with most arterial bifurcations in humans, the carotid bifurcation and ICA are commonly associated with the build up of atherosclerotic plaque or carotid artery disease (CAD). CAD presents as a narrowing, or stenosis, of the vessel that in turn alters blood flow heamodynamics. A representation of which is shown in Figure 1, highlighting the plaque lining the vessel wall and thereby reducing the lumen path for blood to flow. A British Heart Foundation statistics report [1] estimates there to be 152,000 strokes in the UK every year, with approximately 1.2 million people having stroke related morbidity [1]. Ischemic stroke is rapidly becoming one of the leading causes of cardiovascular event-related deaths worldwide, with carotid plaques causing stenosis or occlusion of the vessels accounting for 20% of ischemic strokes [2]. As well as the risk of stroke, CAD is predictive for atherosclerosis within other vascular beds (heart, legs) which all contribute to rates of major adverse cardiovascular events (MACE) [3]. MACE is a commonly used clinical composite endpoint which groups together a variety of undesirable cardiovascular events. It tends to include stroke, myocardial infarction, heart failure and cardiovascular death [4], [5], [6].

One approach to understanding CAD development, and the potential to predict MACE risk, is expected to be closely linked to the local haemodynamics. Carotid artery flow is primarily influenced by local anatomy and upstream flow conditions leading into the vessel. In cases of diseased vessels, the plaque deposition in the lumen can also influence the flow field. The carotid bifurcation angle, the presence and extent of the carotid sinus, tortuosity and diameter are amongst the parameters which significantly impact flow. In certain cases flow separation and flow recirculation can impact adversely, exacerbating local vessel disease, pressure drop and/or flow imbalance. It is often not possible to accurately measure relevant haemodynamic metrics *in vivo*, which lends to the use of computational tools to carry out *in silico* analyses of flow velocities and patterns, wall shear stress (WSS), recirculation, and pressure distributions.

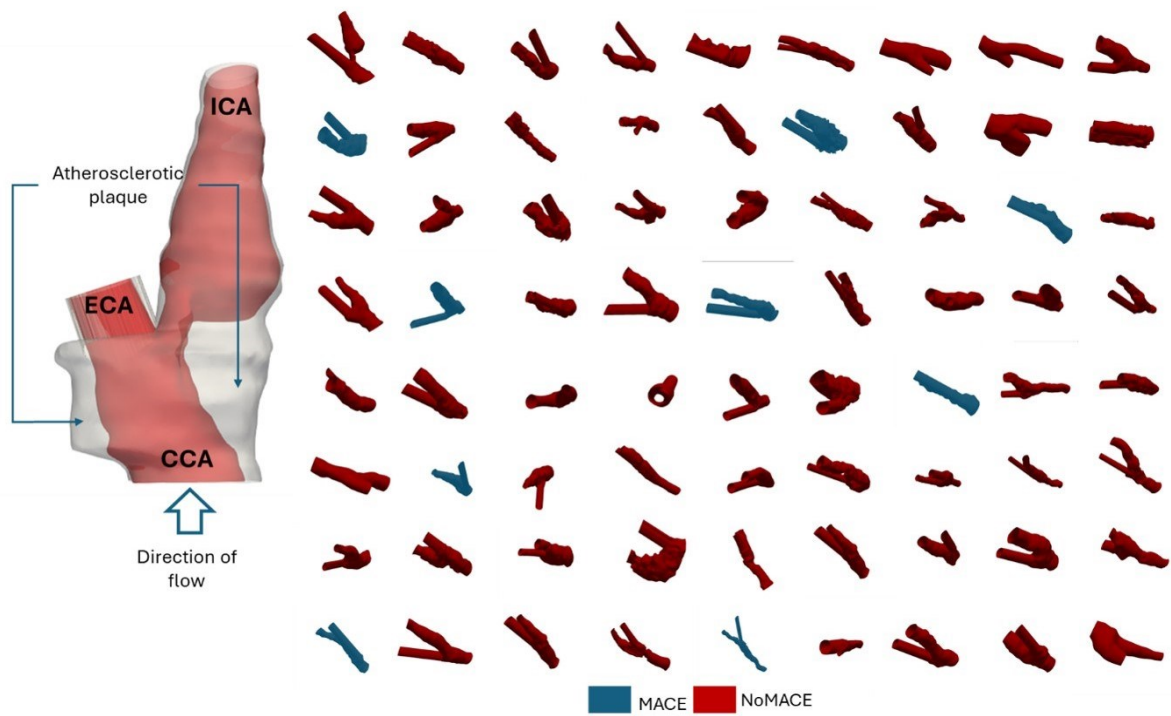

*Figure 1: (Left) Carotid artery with significant plaque lining the walls of the vessel. The region shaded in red represents the actual lumen of the vessel remaining for blood to flow through whilst the unshaded regions represent the plaque lining the vessel wall protruding into the lumen. Flow enters through the common carotid artery (CCA), and bifurcates into the internal carotid artery (ICA) and external carotid artery (ECA).*
*(Right) Carotid artery vessels modelled in this study showing a wide range of geometric variation for each case. Whilst they all have CAD, the cases that suffered from MACE are shaded in blue.*

This study aimed to develop computationally efficient patient-specific models using 3D tomographic ultrasound (tUS) data to examine flow in diseased carotid arteries. The haemodynamic metrics obtained from these models were used to calculate their statistical association with the occurrence of MACE, and their utility for use in computational tools in a clinical setting.

## 2. Methodology

*2.1. Governing equations*

Flow simulations were carried out by solving the governing laws of fluid motion, i.e. the continuity and Navier-Stokes equations. Blood is considered to be a laminar, incompressible, homogeneous and non-Newtonian fluid. This implies that the density of the fluid remains constant and the relative motion between different components of fluid is negligible. The equations for conservation of mass and momentum for incompressible flow are:

$$\nabla \cdot \boldsymbol{u} = 0 \tag{1}$$

$$\frac{\partial \boldsymbol{u}}{\partial t} + \boldsymbol{u} \cdot \nabla \boldsymbol{u} = -\frac{1}{\rho} \nabla p + \nu \nabla^2 \boldsymbol{u} \tag{2}$$

where $\boldsymbol{u}$ is the velocity vector, $t$ is time, $\rho$ is the density of the fluid, $p$ is the pressure and $\nu$ is the kinematic viscosity of the fluid. Equations 1 and 2 govern the mechanics of fluid motion and are solved together when performing haemodynamic modelling.

The shear-thinning nature of blood is considered in this study as low shear rates observed in regions of large recirculation are of importance. Rheological models such as the Cassion model, Power Law model, and Bird-Carreau model are commonly used to describe the nature of blood flow. However, in this investigation, the Casson model has been used as it describes a shear-thinning fluid with a yield stress term that explains blood viscosity behaviour in smaller vessels undergoing recirculation, especially in low-shear regions, and is given as:

$$\mu = \frac{\left(\sqrt{k_0} + \sqrt{k_1 \dot{\gamma}}\right)^2}{\dot{\gamma}} \tag{3}$$

where the dynamic blood viscosity $\mu$ is described in terms of the yield stress $\dot{\gamma}$ and two rheological parameters, $k_0$ ($3.934 \times 10^{-6}$) and $k_1$ ($2.903 \times 10^{-6}$) [7], [8], [9].

*2.2. Image acquisition and geometry reconstruction*

Patient-specific geometries of the carotid artery were reconstructed to include the CCA, carotid bifurcation and ICA. 3D tomographic ultrasound (tUS) data is used in an AI-driven semiautomated software (PIUR tUS Infinity, PIUR imaging, Vienna, Austria) to facilitate the measurement of carotid plaque volume, further details of which have been discussed in previous studies [10], [11]. The 3D tUS data was used to generate standard tessellation language (STL) files, which incorporated the plaque volume lining the vessel walls. These were processed on a computer-aided design software, Meshmixer (v3.5, Autodesk Inc.) to artificially include the ECA based on the orientation and dimensions of the vessel mouth at the bifurcation as they were not actively measured at the 3D tUS stage; they were then iteratively smoothed using the uniform smoothing function on the Meshmixer to reduce the noise on the surface of the STL models and eliminate any sharp edges or vertices which would affect the simulation results. Once suitably generated, the reconstructed geometry was clipped to represent the model inlet at the CCA and two outlets at the ICA and ECA. An example of the reconstruction process using 3D tUS is shown in Figure 2. The region of interest was based on the location of typical atherosclerotic narrowing, namely the carotid bifurcation and proximal ICA (sometimes referred to as the carotid sinus or bulb).

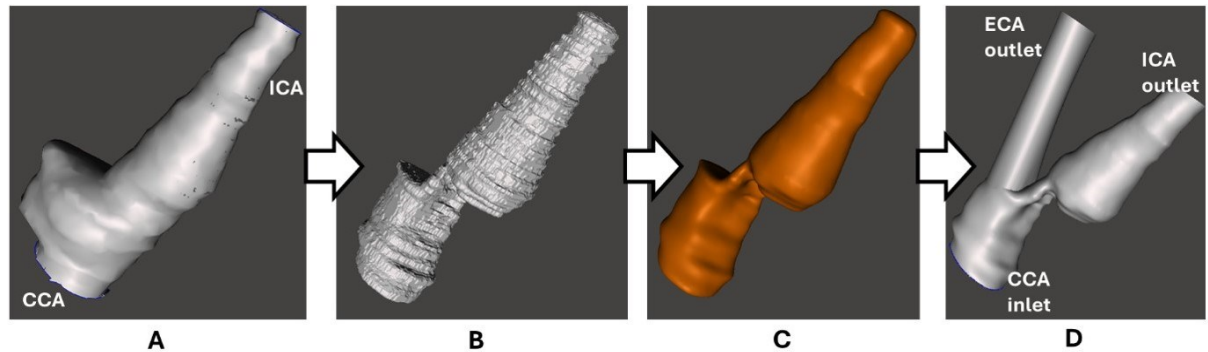

*Figure 2: Reconstruction of anatomically representative 3D geometry from 3D tUS data. A: Overall vessel geometry obtained following the segmentation of the 3D tUS data. B: Actual lumen incorporating plaque lining vessel wall. C: Lumen smoothened to set up fluid domain for simulation. D: ECA manually added as outlet for model to better represent anatomical conditions.*

*2.3. Numerical implementation*

A steady parabolic velocity profile is imposed at the model inlet and with a volumetric flowrate of 5.28 ml.s$^{-1}$, representative of the flowrate [8], under peak flow conditions in the carotid arteries [12]. Varying flowrates lying within a physiological range have been utilised together with different geometries to test the effects of inlet flowrates on haemodynamic metrics, and has been elaborated upon further in the results and discussion section.

The fluid domain is spatially discretised into an unstructured mesh using an automated process on *snappyHexMesh* with the number of cells varying based on the variation of the vessel geometry. A sensitivity test was conducted to establish mesh indepence: the cell size was varied for a representative geometry, thereby varying the total number of cells, and the peak flow velocity was spatially averaged over the entire geometry and compared, and results were compared independent of the mesh when the difference relative to the finer mesh was less than 5%. Geoemtries with smaller and narrower vessels had approximately around 75,000 cells and larger geometries having up to 150,000 cells. This entailed an automated method which creates a base hexahedral mesh enveloping the vessel geometry which then undergoes iterative regional refinement based on the features of the vessel, thus ensuring increased mesh density in regions of interest. The generated meshes then underwent a quality check based on criteria defined in the *snappyHexMesh* script, which performed automatic adjustments and remeshing if necessary. All simulations have been carried out using the finite-volume steady-state solver pimpleFOAM on OpenFOAM (v2212), with pre- and post-processing of data being carried out on ParaView. Spatial and temporal discretisation as achieved used a Gauss linear and second-order implicit backwards scheme respectively. All simulations converged to a residual of 1e-5 with a time-step of 0.001s. The simplified model allowed for efficient simulation times, varying between 51 minutes for smaller geometries to 117 minutes for larger geometries, with an average simulation time of approximately 85 minutes for most geometries. Blood is modelled with a constant density of 1060 kg.m$^{-3}$ and the model inlet at the CCA exhibits largely parabolic flow, comparable to PC-MRI data [13], [14]. The outlets at the ICA and ECA are prescribed as zero-pressure outlets. Walls were assumed rigid with a no-slip boundary condition.

*2.4. Statistical analysis*

Statistical analyses have been carried out by processing data on IBM SPSS Statistics (v28.0.1.1, IBM Inc.). Mean, median and standard deviation have been used to summarise and compare certain haemodynamic metrics between two populations in the dataset. Geometry-averaged metrics of fluid velocity, vorticity, average WSS and peak WSS have been computed from the simulation results. These were then compared between cases that had MACE (9 cases) and those that had not (66 cases), all of which represented a wide range of geometric variation as can be seen in Figure 1. A One-Way ANOVA (analysis of variance) test was carried out to compare the population means between the two sets of data with a 95% confidence interval and statistical significance $\alpha = 0.05$. A linear logistic regression analysis was conducted to determine the association of the aforementioned haemodynamic metrics with MACE. The resultant Receiver Operating Characteristic (ROC) curve would then provide information in terms of the Area Under the Curve (AUC) for the correlation between a chosen variable or combination of variables and the occurrence of cardiovascular events.

## 3. Results

*3.1. Model sensitivity study*

This section focusses on the model sensititivty study carried out to determine the optimal requirements to perform reliable and physiologically representative simulations of flow through carotid arteries.

*3.1.1. Viscosity model sensitivity*

Figure 3 shows a preliminary qualitative comparison of the flow velocities, for one geometry, demonstrates similar patterns in all cases considered. Notable differences are observed in the ECA, where due to the nature of the vasculature, the flow exhibits a certain degree of recirculation. The observed velocity streamlines show minor differences in this region; however, no significant change is observed in the region of interest in the bifurcation and ICA. From a qualitative observation, changing the rheological model does not significantly change the flow pattern.

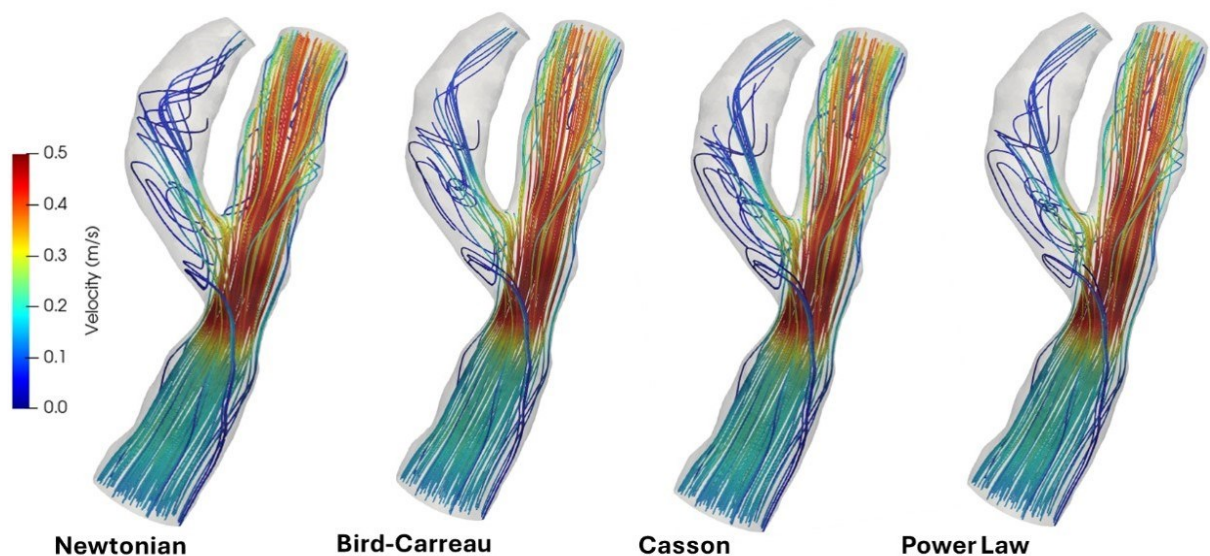

*Figure 3: Velocity streamlines depicting comparable flow patterns observed in the same vessel for different flow viscosity models; Newtonian flow and non-Newtonian flow models: Bird-Carreau, Casson, and Power Law.*

In order to further quantitatively study the sensitivity, the influence of the viscosity on the averaged WSS in the bifurcation and the ICA is considered. As shown in Figure 4, the increase in viscosity averaged over the region of interest is reflected in the WSS being highest in this region among the non-Newtonian models. Further, it is observed that the Casson flow model shows the largest deviation from the WSS observed using Newtonian model. Although the Bird-Carreau and Power Law models exhibited similar trends of average WSS in the same region, yet Casson exhibited the greatest deviation from the Newtonian assumption, which was consistent with the largest increase in flow viscosity. The Casson model was employed for all consequent simulations.

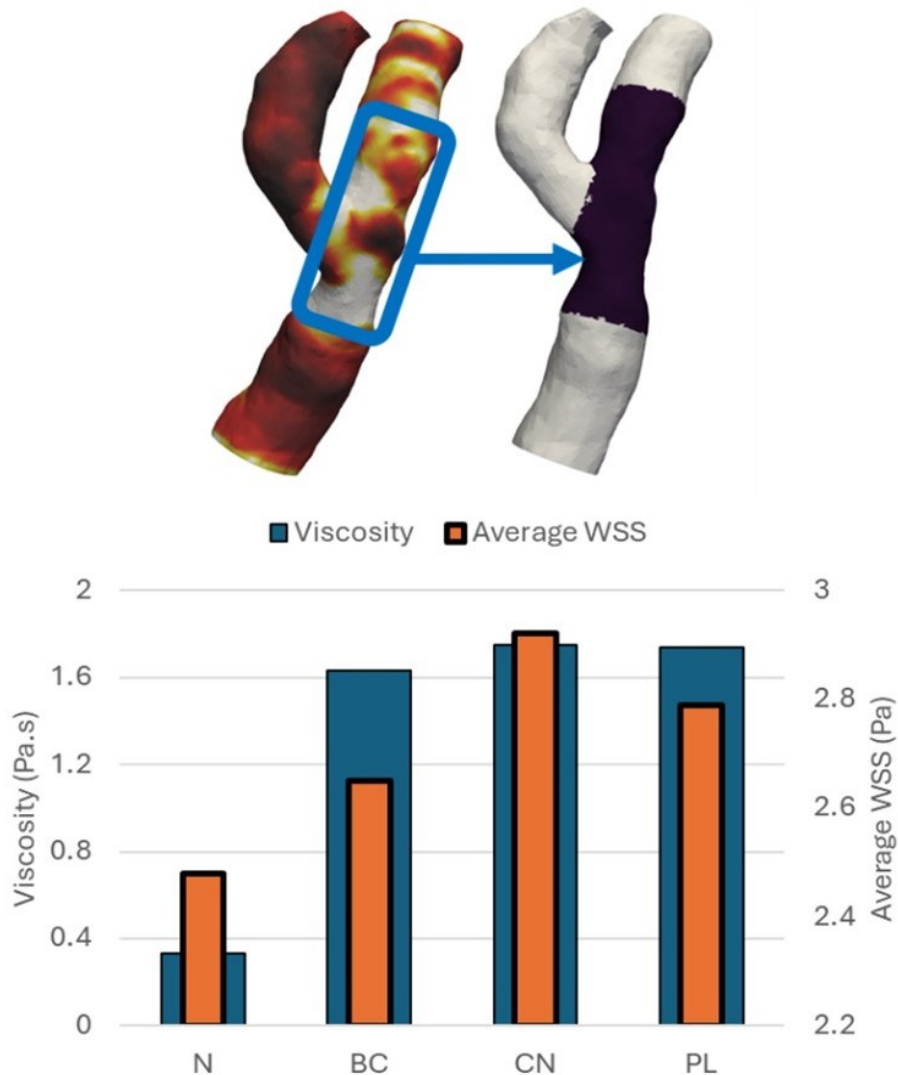

*Figure 4: (Top) Region of interest in which viscosity and WSS was measured, including carotid bifurcation and proximal ICA. (Bottom) Plot showing viscosity and WSS observed in different flow models (NL Newtonian, BC: Bird-Carreau, CN: Casson, PL: Power Law), averages across the region of interest.*

*3.1.2. Inlet flowrate sensitivity*

A range of flowrates falling within a physiologically realistic range have been simulated on three different models, with varying geometric features. The effect of the flowrate would be primarily observed in, and affected by the model inlet. Therefore, the geometries here were chosen to represent different inlet diameters and lengths. Figure 5 shows the three models, Models A, B, and C, that have been used for this comparison. Model A has an approximate length of CCA, from the inlet to the carotid bifurcation, that is comparable to that of C. Model B has a CCA length approximately half that of A and C. However, Models B and C have a similar inlet dimeter compared to that of A. The geometries and their observed peak flow velocities at different flowrates have been shown in Figure 5.

The peak velocities increased with an increase in the inlet flowrates for all three cases. However, Model B, presenting with the shortest CCA length in this case has lower peak velocities compared to the other two. The investigation shows the observable trends in varying the inlet flowrates, with a percentage difference of approximately 9-14% in peak WSS and 6-13% in peak velocity, noted across the range of flowrates for all three geometries.

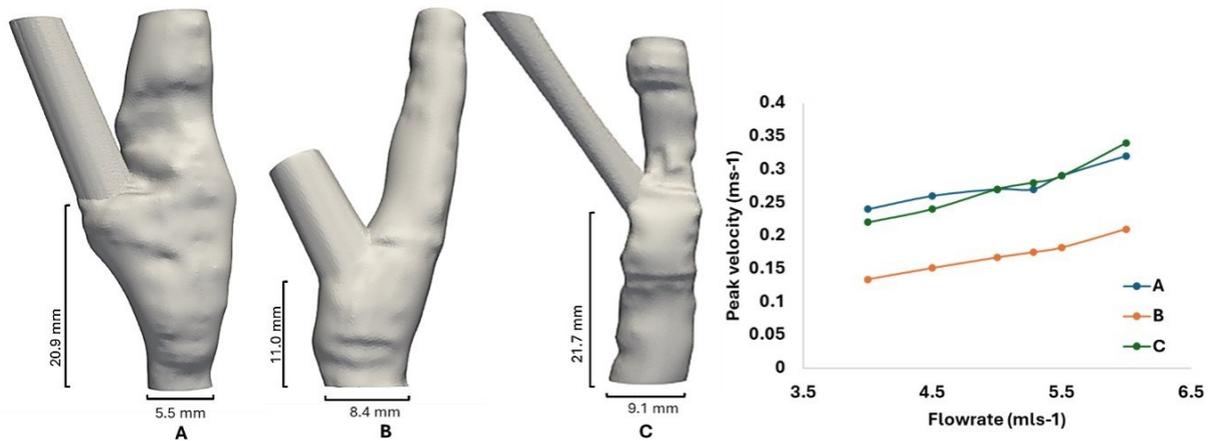

*Figure 5: Models A, B, and C depict a range of inlet diameters and CCA lengths to compare the effect of inlet flowrates on metrics of interest and flow patterns; plot of peak velocity observed in model against flowrate imposed at model inlet.*

*3.2 Statistical analysis for MACE cases*

Figure 6 provides an insight into the flow patterns and spatial distribution of WSS for carotid arteries which have significant plaque lining the lumen wall. There does not appear to be any significantly observable difference between the MACE and NoMACE cases in terms of flow distribution and magnitudes of velocity. However, the MACE cases do exhibit slightly more disturbed flow in the region of interest, i.e. the bifurcation region and ICA, with observable lower velocity recirculating streamlines. Sudden protrusions into the lumen by plaques also give rise to spikes in WSS in the ICA specifically for the MACE cases when compared to the NoMACE cases. It must be noted that the cases shown in Figure 6 only represent a small number of NoMACE cases amongst the dataset that has been modelled, and the aforementioned inference in based on visual inspection.

In Figure 7, the plots show the distribution of all simulated NoMACE and MACE cases for the following geometry-averaged metrics: velocity, vorticity, average WSS, and peak WSS. A One-Way ANOVA test has been conducted to compare the population means for the aforementioned metrics between MACE and NoMACE cases and the p-values obtained are show in Figure 7. This demonstrates that the geometry-averaged vorticity provided the lowest p-value when comparing MACE and NoMACE cases, whilst peak WSS had the highest p-value and thus lowest likelihood of serving as a reliable predictor. Peak WSS, whilst often associated with vascular remodelling, does not show a strong relationship with the likelihood of MACE in this investigation. This could perhaps be attributed to erroneous geomteries resulting in spikes of WSS in certain regions along the vessel wall and this should therefor be considered whilst making inferences. The data reported in Table 1 coveys that the means for all metrics is greater in MACE cases than NoMACE cases.

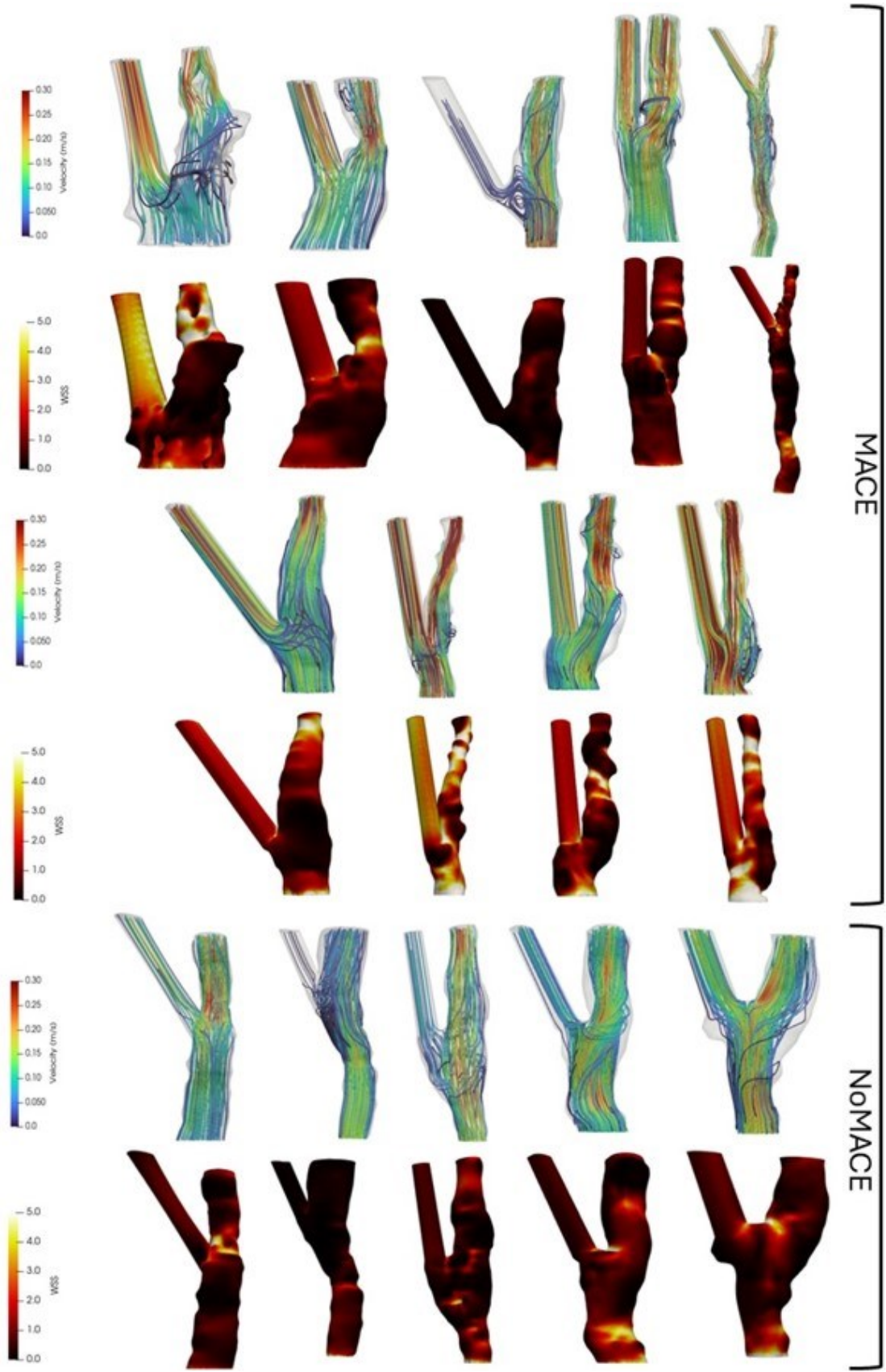

*Figure 6: Instantaneous velocity streamlines and spatial distribution of WSS in carotid arteries that exhibit CAD, with the top two rows showing cases which encountered MACE and the third row with cases that have not encountered MACE.*

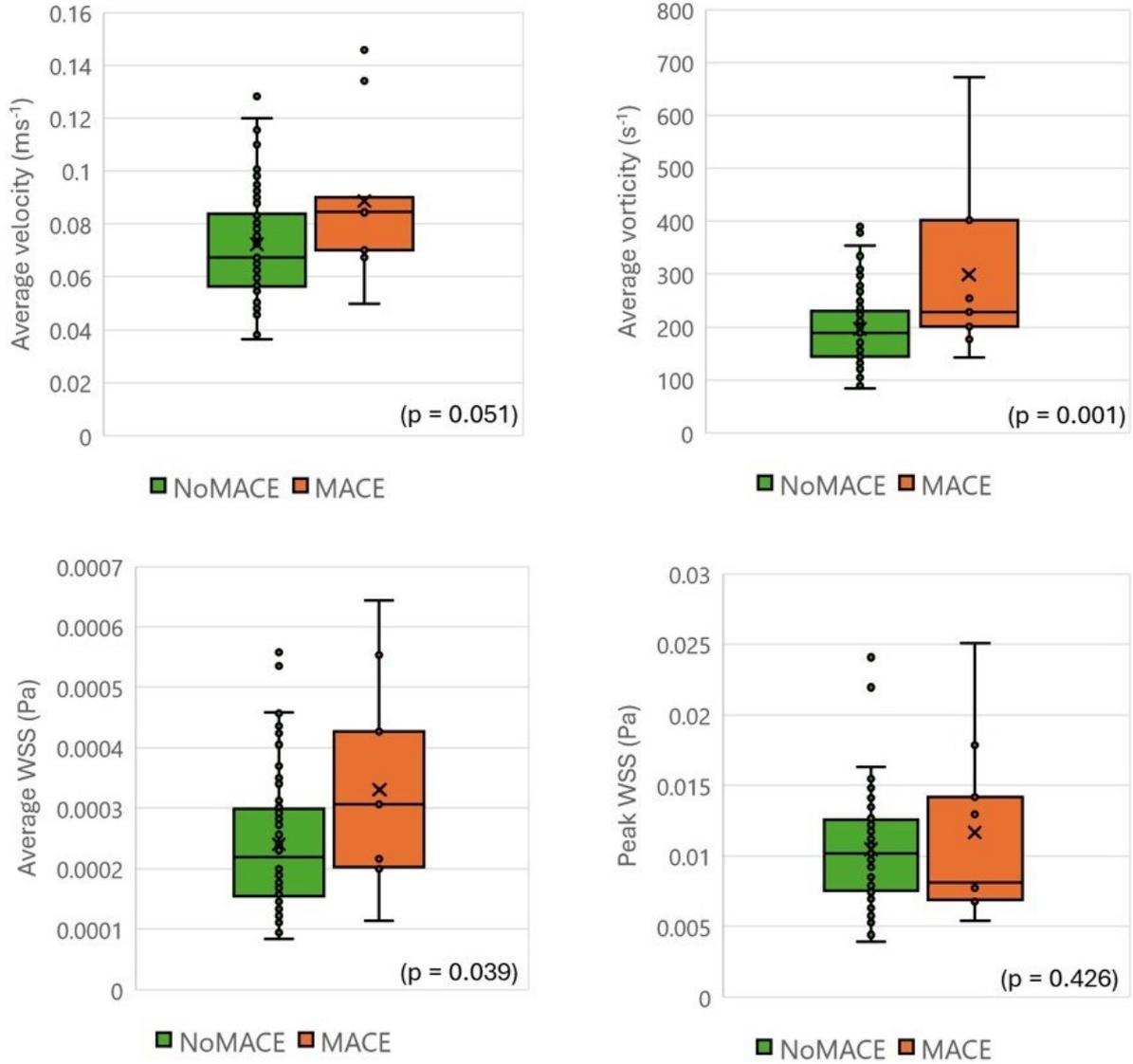

*Figure 7: Plots showing the distribution of data points obtained for MACE and NoMACE cases, including the mean (marked by x in the box plot), median (marked by the horizontal line in each box plot), range (marked by the extended whiskers denoting the maximum and minimum values excluding outliers) and outliers (data points lying beyond the range of the plot) range within the measured metrics. The metrics shown here are geometry-averaged: velocity, vorticity, WSS and peak WSS.*

Table 1: ANOVA values for chosen geometry-averaged haemodynamic metrics across all cases in the study, including MACE and NoMACE cases.

|  | **Average velocity (ms$^{-1}$)** | | **Average vorticity (s$^{-1}$)** | | **Average WSS (mPa)** | | **Peak WSS (mPa)** | |
|---|---|---|---|---|---|---|---|---|
|  | **Mean** | **SD** | **Mean** | **SD** | **Mean** | **SD** | **Mean** | **SD** |
| **NoMACE** | 0.072 | 0.021 | 196.22 | 70.28 | 0.24 | 0.11 | 10.46 | 3.79 |
| **MACE** | 0.088 | 0.031 | 298.73 | 168.36 | 0.33 | 0.17 | 11.65 | 6.51 |

Table 2 further demonstrates the outcome of combining multiple metrics to form an indicator for the occurrence of MACE. Figure 8 shows a ROC curve where all the metrics have been combined in a regression analysis to determine their correlation with MACE. The results from the ANOVA prompted the exclusion of peak WSS from the combined indicator AUC and it showed a negligible decrease from the combined indicator, thereby demonstrating that peak WSS values have low predictive value when it comes to the occurrence of MACE.

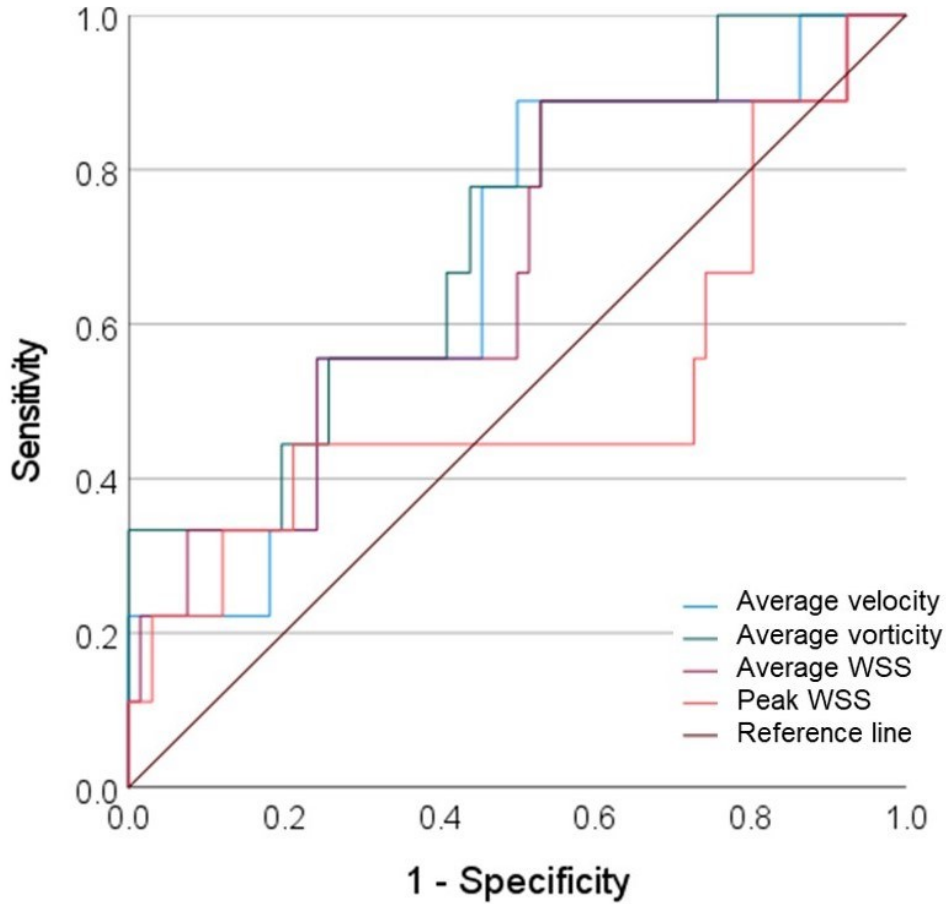

*Figure 8: ROC curve for all the metrics of interest used based on logistic regression analysis, to depict predictability of each metric using Area Under the Curve (AUC) values calculated from plot.*

Table 2: Area Under the Curve values for chosen geometry-averaged haemodynamic metrics across all cases in the study, including MACE and NoMACE cases, following Univariate and Multivariate Logistic Regression Analysis.

|  | Area | Std. error | Asymptotic 95% | |
|---|---|---|---|---|
|  |  |  | Lower bound | Upper bound |
| **Average Velocity** | 0.673 | 0.093 | 0.490 | 0.856 |
| **Average Vorticity** | 0.712 | 0.091 | 0.535 | 0.89 |
| **Average WSS** | 0.662 | 0.102 | 0.463 | 0.861 |
| **Peak WSS** | 0.515 | 0.124 | 0.272 | 0.758 |
| **All metrics combined** | 0.724 | 0.09 | 0.548 | 0.9 |
| **All metrics combined (without Peak WSS)** | 0.722 | 0.089 | 0.531 | 0.913 |

## 4. Discussion

*4.1. Model sensitivity study*

In case of larger vessels like the aorta and its major branches, blood is often modelled under the assumption of being a Newtonian fluid where the shear rates are usually greater than 100 s$^{-1}$ [7], [15], [16]. However, rheological studies of human blood confirm its non-Newtonian and shear-thinning viscous behaviour. Although a Newtonian model is often adequate, this has been compared with different non-Newtonian flow models in order to study the sensitivity of the viscous terms, particularly pertaining to the flow simulation in the carotid. In this regard, a qualitative comparison of the flow profiles and a quantitative study on the influence of the viscosity on the wall shear stress is considered. A comparison is provided between the assumption of Newtonian vs. non-Newtonian (Bird-Carreau, Casson and Power Law) models. Figures 3 and 4 demonstrated there to be no major differences in the flow patterns observed, especially in the bifurcation and ICA. The flow velocity remained largely consistent for the different rheological models. Flow through stenosed vessels, especially associated with atherosclerosis, is associated with large areas of recirculation downstream of the stenosis and high WSS immediately upstream of the narrowing. These phenomena are best represented by modelling blood as a Casson fluid by incorporating the yield stress component of the Casson model expression, which is representative of an initial resistance to flow experienced in low-shear regions [17], [18]. The sensitivity of variation in WSS with changed in flow viscosity is best observed in the Casson model. This work, however, does not consider the time-varying nature flow and limited to the case of steady-state flow representing the peak flow conditions. Therefore, with the peak WSS expected to occur at peak flow conditions, and sensitivity of WSS being important in this study, the Casson model was employed for all consequent simulations.

The flowrate sensitivity study has been carried out in three geometries which represent variations in the CCA inlet which are closely linked to the flow conditions imposed at the model inlet. The flowrate imposed has been obtained from previous studies in literature [12], [19] that have experimentally determined the appropriate flowrate in the CCA. Expectedly, the peak velocities increase with increase in flowrate and thus, inlet velocity, a trend observed in all three cases. It must be noted that this is the peak velocity throughout the geometry, and therefore can be influences by sudden narrowing of the lumen downstream. This investigation only demonstrated the trends in varying the inlet flowrates and exhibited comparable increases in peak flow velocity and peak WSS with increase in the inflow velocity. Altering flowrates presented with an expected trend in flow throughout the vessel; however, there was no significant alteration in haemodynamic metrics of interest and flow patterns observed. Therefore, a physiological average flowrate of 5.28 mls$^{-1}$ [12] was prescribed at the inlet to be suitable for all steady-state simulations performed thereafter for consistency and reliability.

*4.2. Statistical analysis for MACE cases*

The incidence of MACE in patients can be associated with haemodynamic changes occurring due to the presence of atherosclerotic plaque lining the arterial wall [3]. CAD results in a higher likelihood of individuals experiencing more serious CVE or MACE. The intention in numerous investigations remains to obtain decisive biomarkers or metrics than be used to predict the likelihood of MACE in patients, especially those who are at high risk with CAD. Despite their not being any significant differences in the pattern of flow, plaques that protrude into the vessel lumen could result in greater degrees of recirculation in flow downstream of the plaque, and spikes in WSS in these regions. This can be observed in the cases shown in Figure 6. The haemodynamic metrics that have the greatest influence on cardiovascular events, and those best suited to serve as indicators for disease have been studied extensively throughout the years and continue to be an area of interest. Patterns of flow, especially sudden acceleration in flow due to a stenosis, and consequently WSS are indices that are commonly associated with disease initiation and progression. This lends to the inclusion of flow velocity. Vorticity in the fluid field gives the measure of rotation at any point in the fluid and was used here as a means to measure the recirculation observed throughout the fluid domain. WSS is commonly associated with atherosclerotic disease progression and can also be a precursor to thrombus formation and therefore the peak and mean values of WSS were included here.

The regression analysis provided the outcome of individual metrics as well as a combination of metrics to form an indicator for the occurrence of MACE. The geometry-averaged values of velocity, vorticity, and WSS were taken as an average across the total computational domain, i.e. the entire vessel geometry. This involves the region of interest, i.e. the CCA leading into the carotid bifurcation, the ECA, the ICA, including the carotid sinus region. The active region of interest, if further reduced to include only the bifurcation and carotid sinus, would reduce the region that the metrics are being averaged over in an attempt to increase result accuracy; however, this would likely result in neglecting part s of the vessel that are significantly plaqued beyond this. Therefore, reducing the region of interest could perhaps increase the AUC, whilst the overall trend remains agreeable with the current method of analysis. Peak WSS demonstrated the least strong relationship with the incidence of MACE, with its inclusion in the ROC curve providing negligible difference to the outcome. It should be noted that vorticity provided a relatively high AUC, thereby confirming the results of the ANOVA where it showed strong association with MACE occurrence. However, the AUC is not high enough for it be considered a reliable predictor on its own. Vorticity being reflective the recirculation observed in the vasculature could be an indication that MACE is associated with recirculating flow induced due to plaque. This is also exhibited in Figure 6, where all MACE cases tended to show regions of non-uniform and recirculating flow in the bifurcation and ICA. The use of computationally obtained haemodynamic indicators to reliably predict the occurrence of disease, in this case more generally MACE, would serve significant clinical benefits. The likelihood of obtaining a sole haemodynamic predictor is low, as cardiovascular disease is influenced by a myriad of factors. A combination of metrics to produce a reliable predictor is a reasonable goal. In this investigation, geometry-averaged vorticity provides reasonable AUC, with a combination of metrics including vorticity, flow velocity and WSS providing a similar AUC. This complies with the comparisons of population means, which showed vorticity measurements to be most significantly different between MACE and NoMACE populations, with peak WSS having the least influence. The distribution of WSS however, is often associated with plaque development [20], [21], [22] and could serve as a predisposition to MACE as has been reported in other studies [23], [24], [25]. Therefore, the absolute magnitudes of WSS potentially have less of an influence on plaque-induced cardiac events, and the spatial distribution of the WSS and the sudden differences in magnitude could have greater influence.

## 4. Summary and future work

Understanding the haemodynamics of carotid artery flow and the influence of plaque on the localised haemodynamics remains of significance to allow clinicians to better tackle CAD and reduce the risk of associated CVD. This study provides a computationally efficient method for modelling flow through the carotid artery. The simplified model allowed for computational efficiency with an average simulation time of 85 minutes with an additional 20 minutes of pre-processing the geometry. 3D tUS data has been utilised to faithfully reconstruct the arterial lumen of the carotid vessels taking into account the presence of atherosclerotic plaque lining the inner walls of the vessels, thereby representing the actual path of flow experienced by fluid in these heavily plaqued vessels. A steady-state analysis with non-Newtonian flow through the vessel allowed for a simplified approach to modelling flow through the carotid vessels. Statistical analyses of relevant metrics showed the association of different metrics with the occurrence of MACE, and the reliability of them being used as indicators in a predictive tool.

The dataset processed for this study encompasses a broad range of vasculature that present with highly varied geometric features and a varying degree of plaque lining the lumen wall. It includes symptomatic patients, some of whom have experienced MACE. The rich range of geometric data and CFD information now available specific to each case makes this a viable training dataset for generating predictive models for CVD. The haemodynamic metrics of interest from a clinical standpoint are often largely dictated by geometric variation of the vasculature, given that there are negligible comorbidities to account for. The training dataset can be used to develop predictive tools which would interpret the data such that it can determine haemodynamic metrics of interest upon geometric features of the vasculature.

Prediction and assessment of cardiovascular events remains a significant area of interest for clinicians. The goal is to develop personalised diagnostic and clinical tools utilising *in silico* methods to tackle and prevent CVE. In this study, the association of haemodynamic metrics with the occurrence of MACE has been investigated, and it lays the foundation for utilising a combination of metrics to serve as an indicator in the development of predictive models for cardiovascular medicine.

*Acknowledgements*: Joao Carriera, Tomorrow Cardiovascular Limited, for image analysis. This project was funded via the Confidence for Translation award from Translation Manchester, which funded Dr Sampad Sengupta. Prof. Frank L. Bowling and Dr Jonathan Ghosh, Manchester Academic Vascular Research & Innovation Centre, for funding acquisition. The 3D tomographic ultrasound scans reused in this project were captured during a personal award to Dr Steven K Rogers via the UK National Institute for Health Research (NIHR) under grant agreement number 301209.

*Credit authorship contribution statement*:
Sampad Sengupta: Methodology, Investigation, Software, Validation, Formal analysis, Visualisation, Resources, Data Curation, Writing - Original draft, Editing;  Emily Manchester: Supervision, Formal analysis, Funding acquisition, Writing - Editing;  Jie Wang: Software, Validation;  Ajay B Harish: Conceptualization, Methodology, Investigation, Resources, Supervision, Project administration, Funding acquisition, Writing - Editing;  Alistair Revell: Conceptualization, Methodology, Investigation, Resources, Funding acquisition, Supervision, Writing - Editing;  Steven Rogers: Conceptualization, Methodology, Investigation, Resources, , Funding acquisition, Data Curation, Supervision, Project administration, Writing – Editing, Approval for submission.

**Abbreviations**

| | |
|---|---|
| **AUC** | Area under the curve |
| **CAD** | Carotid artery disease |
| **CCA** | Common carotid artery |
| **ECA** | External carotid artery |
| **ICA** | Internal carotid artery |
| **MACE** | Major adverse cardiovascular event |
| **ROC** | Receiver operating characteristic |
| **STL** | Standard tessellation language |
| **tUS** | Tomographic ultrasound |
| **WSS** | Wall shear stress |